# Generation of 1.5 μm discrete frequency entangled two-photon state in polarization maintaining fibers

Qiang Zhou,[1, 2, 3] Wei Zhang,[1,*] Chenzhi Yuan,[1] Yidong Huang,[1] and Jiangde Peng[1]

[1] Tsinghua National Laboratory for Information Science and Technology, Department of Electronic Engineering, Tsinghua University, Beijing, 100084, P. R. China
[2] School of Optoelectronic Information, University of Electronic Science and Technology of China, Chengdu, 610054, P. R. China
* *zwei@tsinghua.edu.cn;*
[3] *betterchou@gmail.com.*

In this Letter, the generation of 1.5 μm discrete frequency entangled two-photon state is realized based on a piece of commercial polarization maintaining fiber (PMF). It is connected with a polarization beam splitter to realize a modified Sagnac fiber loop (MSFL). Correlated two-photon states are generated through spontaneous four wave-mixing process along the two propagation directions of the MSFL, and output from the MSFL with orthogonal polarizations. The quantum interference of them is realized through a 45° polarization collimation between polarization axes of PMFs inside and outside the MSFL, while the phase difference of them is controlled by the polarization state of the pump light. The frequency entangled property of the two-photon state is demonstrated by a spatial quantum beating experiment with a fringe visibility of (98.2±1.3)%, without subtracting the accidental coincidence counts. The proposed scheme generates 1.5 μm discrete frequency entangled two-photon state in a polarization maintaining way, which is desired in practical quantum light sources.

Discrete frequency entangled two-photon state is a potential resource for quantum information applications, such as quantum networks with distributed stationary quantum nodes [1, 2], quantum communication and cryptography with higher capacity [3-5], improved quantum communication in noisy quantum channels [6], enhanced quantum clock synchronization [7, 8], and nonlocal dispersion cancellation in quantum interferometry [9]. One way to generate the discrete frequency entangled two-photon state is based on the quantum interference of two frequency non-degenerate correlated two-photon states, which has been demonstrated in nonlinear crystals [10], dispersion shifted fibers [11], and silicon waveguides [12] through the second-order or the third-order spontaneous nonlinear optical processes. Among them, the fiber based scheme holds the advantages of compatible with current technologies in optical fiber communication and is easy to develop practical quantum light sources at telecom band [11, 13-19]. The two-photon state is efficiently generated in well-defined single fiber modes through spontaneous four wave-mixing (SFWM) and can be conveniently manipulated by linear optical devices to develop complex quantum information functions with high efficiency due to the low loss properties of optical fibers and fiber based devices. However, polarization fluctuations in optical fibers may impact the long-term stability of the fiber based scheme in Ref. 11, in which the two correlated two-photon states are generated in a piece of dispersion shifted fiber in a Sagnac fiber loop, and the quantum interference of them is controlled by a fiber polarization controller in the loop to collimate the polarization states of them. In this Letter, we propose and demonstrate a scheme of discrete frequency entangled two-photon state generation based on a piece of commercial polarization maintaining fibers (PMFs). It is connected with a polarization beam splitter (PBS) to realize a modified Sagnac fiber loop (MSFL). Correlated two-photon states are generated through SFWM along the two counter propagation directions of the MSFL, respectively, and output from it with orthogonal polarizations. The quantum interference of them is realized through a 45° polarization collimation between polarization axes of the PMFs inside and outside the MSFL, while the phase difference of them is controlled by the polarization state of pump light. This scheme generates 1.5 μm discrete frequency entangled two-photon states in a polarization maintaining way, which is desired in practical quantum light sources.

Figure 1 shows the experiment setup for this scheme. The pulsed pump light is provided by a gain-switched semiconductor laser diode. The peak power of pump pulses is amplified by an erbium doped fiber amplifier (EDFA). The central wavelength, pulse width and repetition rate of pump pulses are 1552.52 nm, 19 picoseconds and 4 MHz, respectively. A variable optical attenuator is used to control the level of pump power, which is monitored by a 50/50 fiber coupler and a power meter (PM) in the experiment. A manual polarization controller fixed on the fiber bench (FB1, Thorlabs Inc., PC-FFB-1550) is used to adjust the polarization state of the pump light. A dense wavelength division multiplexer (DWDM) at 1555.52 nm (ITU-C31, side-band rejection >120 dB) is used to suppress the amplified spontaneous emission (ASE) noise of the EDFA, ensuring the effective detection of the generated photon pairs. In the experiment, the DWDM is also used to inject pump light into the MSFL from transmission port to common port (T to C) and extract generated two-photon states exiting from the MSFL (C to R). All the devices are connected by PMF pigtails.

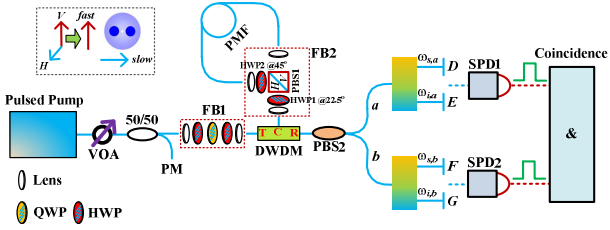

Fig. 1. (Color online) Experiment setup. VOA: Variable Optical Attenuator; PM: Power Meter; FB: Fiber Bench; DWDM: Dense Wavelength Division Multiplexer; PBS: Polarization Beam Splitter; SPD: Single Photon Detector; QWP: Quarter-Wave Plate; HWP: Half-Wave Plate; 50/50: 50/50 Fiber Coupler.

The MSFL consists of a PBS (PBS1) installed on a fiber bench (FB2, Thorlabs Inc., PFS-FFT-1X2-1550) with two rotatable half-wave plates (HWP1 and 2) and a piece of commercial PMF (50 meters in length, Fujikura Ltd.), which is cooled by liquid nitrogen to reduce Raman noise photons [14]. The PBS1 has two polarization axes denoted by $H$ and $V$, while the PMF and PMF pigtails of the devices have the fast and slow polarization axes, which are collimated with $V$ and $H$, as shown in Fig. 1. In order to realize discrete frequency entangled two-photon state generation, polarization direction rotations need be introduced, which is achieved by setting HWP1 and HWP2 at 22.5° and 45° with respect to the $V$ axis, respectively. Let elliptically polarized pulsed pump light with its long elliptical axis paralleling with $V$ axis be injected into the MSFL, after passing through the HWP1 and PBS1, the pump light is equally split into $H$ and $V$ components with a phase difference $\varphi_p$, which is calculated by [15],

$$\varphi_p = \pm 2 \times \operatorname{atan}\left(10^{-R_{per}/10}\right) \quad (1)$$

Where, $R_{per}$ is the polarization extinction ratio of pump light; + and − correspond to right- and left- handed elliptically polarized light, respectively. And then the $H$ component is rotated into $V$ component after passing through the HWP2. Hence the two pump components inject into the commercial PMF paralleling with the same polarization axis. One propagates in the clock wise (CW) direction and the other propagates in the counter clock wise (CCW) direction. The two pump components would generate correlated two-photon states through SFWM processes and output from the MSFL with orthogonal polarizations. Considering that the correlated photon pair generation rates are far smaller than 1 pair per pulse, two-photon states output from PBS1 can be expressed as,

$$\psi_0 = \frac{1}{\sqrt{2}}\left(|\omega_s \omega_i\rangle_{H,CCW} + e^{i2\varphi_p}|\omega_s \omega_i\rangle_{V,CW}\right) \quad (2)$$

Where, $\omega_i$ and $\omega_s$ are the angular frequencies of idler and signal photons, respectively. The obtained two-photon states pass through HWP1, and then enter into the DWDM from the common port (C), finally output from its reflection port (R). Another PBS with PMF pigtails (PBS2) connected to the reflection port (R) of DWDM. Thanks to the 22.5° setting of the HWP1, polarization axes of PMFs outside the MSFL, i.e. PMF pigtails of the DWDM and the PBS2, have a 45° offset with respect to the polarization axes inside the MSFL, which realizes the quantum interference between the two correlated two-photon states paralleling along $H$ and $V$ polarization directions output from the MSFL. The two output ports of the PBS2 is denoted by port a and port b, respectively. Due to the quantum interference, the two-photon states output from PBS2 is,

$$\psi_1 = \frac{1}{2}\left(\left(1+e^{i2\varphi_p}\right)\psi_b + \left(1-e^{i2\varphi_p}\right)\psi_{ab}\right)$$
$$\psi_b = \frac{1}{\sqrt{2}}\left(|\omega_s\rangle_a |\omega_i\rangle_a + |\omega_s\rangle_b |\omega_i\rangle_b\right) \quad (3)$$
$$\psi_{ab} = \frac{1}{\sqrt{2}}\left(|\omega_s\rangle_a |\omega_i\rangle_b + |\omega_s\rangle_b |\omega_i\rangle_a\right)$$

Where, $|\omega_{s,i}\rangle_a$ and $|\omega_{s,i}\rangle_b$ are the photon states with signal or idler photon outputs from port a and port b, respectively. Hence, $\psi_b$ is the spatial bunched path-entangled two-photon state, i.e. the two photons output from the same port simultaneously; $\psi_{ab}$ is the spatial anti-bunched path-entangled two-photon state, i.e. the two photons output from different ports simultaneously. From Eq. (3), it can be seen that the two-photon state exiting from PBS2 is a superposition of the spatial bunched and anti-bunched two-photon states. The output possibilities of them can be expressed as,

$$P_{bu} = 0.5 \times \left(1 + \cos(2\varphi_p)\right)$$
$$P_{ab} = 0.5 \times \left(1 - \cos(2\varphi_p)\right) \quad (4)$$

Where, $P_{bu}$ and $P_{ab}$ are the output possibilities of $\psi_b$ and $\psi_{ab}$, respectively. It can be seen that the output possibilities of the two two-photon states are varied with $\varphi_p$, which is due to the quantum interference of the two correlated two-photon states. Pure spatial bunched or anti-bunched path-entangled two-photon states can be obtained under $\varphi_p$ of $k\pi$ or $(k+0.5)\pi$, where $k$ is an integer. The spatial anti-bunched two-photon states is the discrete frequency entangled two-photon state due to the frequency non-degenerate property of the generated correlated two-photon state in the commercial PMF. It is worth noting that the fiber bench devices are used in our experiment demonstration, however these devices can be substituted by polarization maintaining fiber devices with very small birefringence walk-off [16, 17].

To measure the output possibilities of the spatial bunched and anti-bunched two-photon states under different $\varphi_p$. The exiting photons from PBS2 are sent into two filtering and splitting modules with insertion losses of 0.8 dB. Signal and idler photons satisfying the energy conservation are selected by the two filtering and splitting modules with a pump light isolation of 120 dB. The selected wavelengths of signal and idler photons are centered at 1555.75 nm and 1549.32 nm respectively, with a -3 dB spectral width of 2π×32 GHz. The photon states exit from ports $D$, $E$, $F$ and $G$ correspond to $|\omega_s\rangle_a$, $|\omega_i\rangle_a$, $|\omega_s\rangle_b$, and $|\omega_i\rangle_b$ photon states in Eq. (2), respectively. The spatial bunched and anti-bunched two-photon state can be distinguished by coincidence measurement with different combinations of the output ports. To measure the spatial bunched two-photon states, ports $D$ and $E$ (or $F$ and $G$) are connected to two single-photon detectors (SPD1 and 2, ID201, ID Quantique), respectively; while for spatial anti-bunched two-photon states, ports $D$ and $G$ (or $E$ and $F$) are connected to SPD1 and 2, respectively. The two SPDs are operated in gated Geiger mode with a 2.5 ns detection window, triggered by residual pump light detected by a photon detector. The detection efficiencies of SPD1 and 2 are 21.8% and 22.6%, respectively. The dark count rates are $5.82 \times 10^{-5}$ and $4.60 \times 10^{-5}$ per gate, respectively. The output electronic signals of the two SPDs are sent into a coincidence circuit, and the coincidence counts are recorded.

From Eq. (1) and (4), it can be seen that $\varphi_p$ is different for left- and right- handed elliptically polarized light with the same $R_{per}$, however output possibilities of the two two-photon states are the same for both the left- and right- handed elliptically polarized lights with the same $R_{per}$. Hence, in the experiment the non-classical two-photon interference (TPI) resulting from the quantum interference is measured under different $R_{per}$, while the long elliptical axis of pump light is set to 0° respecting to $V$ axis without regarding the left- or right- handed polarization property of the pump light. The polarization state of pump light is monitored by a polarization analyzer (Santec PEM-320) at the end of PMF pigtail at the common (C) port of DWDM.

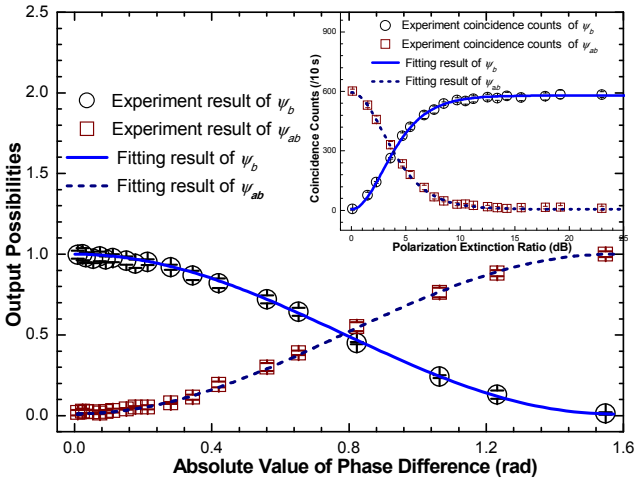

Fig. 2. (Color online) The experiment results of the non-classical TPI. The main figure is the output possibilities. Circles and squares are the experiment results of the spatial bunched and anti-bunched two-photon states, respectively. Solid and dashed lines are the fitting curves of them. The inset gives the measured coincidence counts under different polarization extinction ratio of pump light.

Figure 2 shows the experiment results of the non-classical TPI. In the experiment, ports $D$ and $E$ are connected with SPD1 and 2 for spatial bunched two-photon state, while ports $D$ and $G$ for spatial anti-bunched two-photon state. The inset of Fig. 2 gives the measured coincidence counts under different $R_{per}$. The output possibilities of the two two-photon states are obtained by normalizing the coincidence counts with the sum of them and shown in the main figure of Fig. 2, where $\varphi_p$ is calculated by Eq. (1). In these figures, circles and squares are the experiment results of the spatial bunched and anti-bunched two-photon states, respectively. Solid and dashed lines are the fitting curve of them following Eq. (1) and (4). Both fitting curves show a non-classical TPI visibility of (98.0±1.2)%, without subtracting accidental coincidence counts ((99.1±1.2)% if accidental coincidence counts are subtracted).

In Fig. 2, it can be seen that pure spatial bunched or anti-bunched path-entangled two-photon states can be prepared with a linearly or circularly polarized pump light injected into the MSFL, respectively. Actually, the spatial anti-bunched path-entangled two-photon state is just the frequency entangled two-photon state. By adjusting the polarization controller installed in FB1, a circularly polarized pump light with a $R_{per}$ of 0.1 dB is obtained. Under this condition the discrete frequency entangled two-photon state is prepared at the output of PBS2. The frequency entanglement of the two-photon state is demonstrated by an experiment of spatial quantum beating. Figure 3(a) shows the experiment setup. The photons output from port a and port b input into a 50/50 fiber coupler with a relative arrival time delay of $\Delta\tau$, which is controlled by a variable delay line (VDL, MDL-002, General Photonics Corp.). PC1 and 2 are used to ensure that the input photons of 50/50 coupler are in identical polarization state. At the output ports of the 50/50 coupler, photons pass through signal and idler filters and then are detected by SPD1 and 2, respectively. The outputs of two SPDs are sent into coincidence circuit to obtain coincidence counts. Normalized coincidence counts can be expressed as [11, 20],

$$P_{co} \propto 1 - V_0 \xi(\sigma \times \Delta\tau) \cos(|\nu_i - \nu_s| \times \Delta\tau)$$
$$\xi(\sigma \times \Delta\tau) = \text{sinc}(\sigma \times \Delta\tau) \quad (5)$$

Where, $V_0$ is the visibility of spatial quantum beating. $\xi(\sigma \times \Delta\tau)$ is related to the spectral properties of signal and idler side filters, which is $\text{sinc}(\sigma \times \Delta\tau)$ if the transmission property of filters for signal and idler photons have Gaussian profiles. $\sigma$ is the angular frequency bandwidth of the filters, which is $\sigma = 2\pi \times 32 \times 10^9$ Hz in the experiment; $\nu_{i,s}$ are the frequencies of idler and signal photons; the frequency spacing of them is 800 GHz in the experiment. Figure 3(b) shows the experiment result of the spatial quantum beating, under a photon pair generation rate of about 0.01/pulse with an average pump power level of 12.6 μW. Circles are experiment results, in which accidental coincidence counts are not subtracted. The solid line is fitting curve according to Eq. (5), showing a visibility of (98.2±1.3)%, without subtracting accidental coincidence counts ((99.2±1.1)% if the accidental coincidence counts are subtracted). The period of spatial quantum beating fringe is 1.25 ps, i.e. a period of 375 μm in length, which is determined by the frequency spacing between the idler and signal photons. It can be seen that the experiment result agrees well with the predication of Eq. (5), demonstrating the frequency entanglement of the prepared two-photon state.

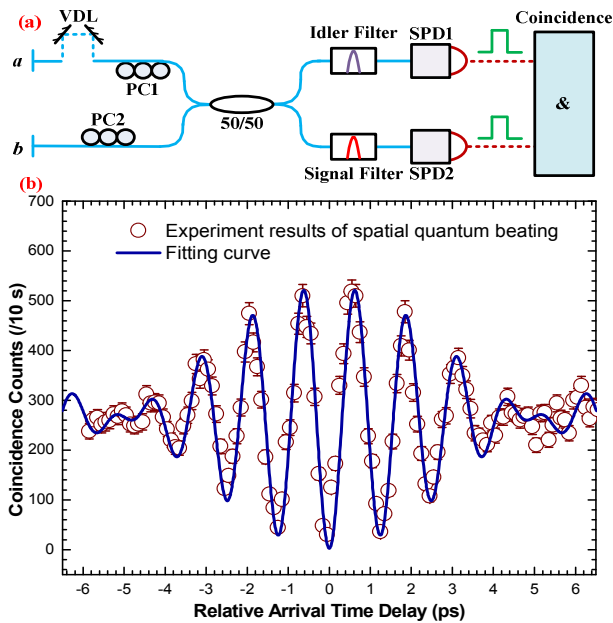

Fig. 3. (Color online) (a) Setup for measuring spatial quantum beating; VDL: Variable Delay Line; 50/50: 50/50 Coupler; PC: Polarization Controller. (b) The spatial quantum beating of discrete frequency entangled two-photon state; Circles are experiment results; solid line is the fitting curve with a fitting visibility of (98.2±1.3)%, without subtracting accidental coincidence counts.

In summary, in this paper the generation of 1.5 μm discrete frequency entangled two-photon state has been realized based on a piece of commercial PMF in a MSFL. Correlated two-photon states generate along the two propagation directions of the MSFL, respectively, and output from it with orthogonal polarizations. The quantum interference of them is realized through a 45° polarization collimation between polarization axes of PMFs inside and outside the MSFL, while the phase difference of the two correlated two-photon state is controlled by the polarization state of pump light. A non-classical TPI was experimentally measured with a fitting visibility of (98.0±1.2)%, without subtracting accidental coincidence counts. Discrete frequency entangled two-photon states were obtained at the output ports of the PBS2, while a circularly polarized pump light is injected into the MSFL. The frequency entangled property was measured by a spatial quantum beating experiment with a fitting visibility of (98.2±1.3)%, under a photon pair generation rate of 0.01/pulse. The experimental results demonstrate that this scheme generates 1.5 μm discrete frequency entangled two-photon state in a polarization maintaining way, which is desired in developing practical quantum light sources for quantum information.

This work is supported by 973 Programs of China under Contract Nos. 2010CB327606 and 2011CBA00303, China Postdoctoral Science Foundation, Basic Research Foundation of Tsinghua National Laboratory for Information Science and Technology (TNList), Research Startup Funds of University of Electronic Science and Technology of China under Contract No. Y02002010501062.